\documentstyle[sprocl]{article}

\input{psfig}

\newcommand{\beq}{\begin{equation}}
\newcommand{\eeq}{\end{equation}}
\newcommand{\bea}{\begin{eqnarray}}
\newcommand{\eea}{\end{eqnarray}}

\newcommand{\tS}{\tilde{S}}
\renewcommand{\P}{{\cal P}}
\newcommand{\tU}{\tilde{U}}

\renewcommand{\ni}{\noindent}

\renewcommand{\d}{\delta}

\renewcommand{\b}{\beta}

\newcommand{\n}{\nu}
\newcommand{\m}{\mu}

\newcommand{\s}{\sigma}

\newcommand{\oh}{\frac{1}{2}}

\newcommand{\dg}{\dagger}
\newcommand{\non}{\nonumber}

\newcommand{\rf}[1]{(\ref{#1})}
\newcommand{\ra}{\rightarrow}

\begin{document}

\title{Center Vortices at Strong Couplings and All Couplings
\footnote{Talk presented at {\sl Confinement 2000}, Osaka,
Japan, March 7-10 2000.}}

\author{J. Greensite}
\address{Physics Dept., San Francisco State University,
San Francisco CA 94132 USA}

\maketitle

\begin{abstract}
Motivations for the center vortex theory of confinement are discussed.
In particular, it is noted that the abelian dual Meissner effect,
which is the signature of dual superconductivity, cannot adequately
describe the confining force at large distance scales.  A long-range
effective action is derived from strong-coupling lattice gauge theory
in D=3 dimensions, and it is shown that center vortices emerge as the
stable saddlepoints of this action.  Thus, in the case of
strong couplings, the vortex picture is arrived at analytically.
I also respond briefly to a recent criticism regarding maximal center 
gauge.
\end{abstract}

    In this talk I would like to present some recent work,
done in collaboration with Manfried Faber and \v{S}tefan Olejn{\'\i}k,
concerning center vortices in strong coupling lattice gauge theory.  I
will also touch on results obtained in collaboration with
Jan Ambj{\o}rn and Joel Giedt,~\cite{j3} and with Faber, Olejn{\'\i}k,
and Roman Bertle.~\cite{coming}
  
\section{Why Center Vortices?}

   We begin with a simple question: If confinement is defined by the
Wilson area-law criterion, then what charge is actually confined in an
SU(N) gauge theory?  The first answer that comes to mind is that
confined charge is just the SU(N) color charge.  But this answer can't
be quite right, at least according to the Wilson criterion, because
not all color charges are confined in this sense.  For example, due to color
screening, there exists no asymptotic linear potential between heavy
charges in the adjoint representation.  A second possible answer,
motivated by dual-superconductor models, is that abelian electric
charge (identified by abelian projection $SU(N) \ra U(1)^{N-1}$) is
the charge that is confined.  But this doesn't work either, since not
all electric charges are confined.  In, e.g., SU(2) lattice gauge
theory, there is no asymptotic linear potential between charges of
$q=\pm 2$ multiples of the elementary charge.  Finally, consider
N-ality, i.e.\ the charge associated with the $Z_N$ subgroup of SU(N).
It is well known that in SU(N) gauge theory, only color charges with
non-zero N-ality are confined, and the asymptotic string tension
depends only on the N-ality of the color charge representation.
Thus we conclude that it is really $Z_N$ charge which is confined in
non-abelian gauge theories.

   This simple fact is interesting, because the type of charge confined
is an indication of the type of field configuration which does the confining.
Of course, the N-ality dependence of the QCD string tension is
no mystery; it is simply due to color screening of higher charge 
representations by gluons.  We have an intuitive picture of one or more 
gluons bound to a static charge.  On the other hand, Wilson loops
can also be interpreted as a probe of vacuum fluctuations in the absence
of external sources (think of evaluating spacelike loops in the Hamiltonian
formulation), and
it is generally assumed that loop observables are disordered by certain 
large-scale topological configurations.  If that is the case,
then such configurations must have the very non-trivial property that 
N-ality $=0$ loops are somehow \emph{not} disordered, and that the induced 
string tension in general depends only on N-ality.  The only known 
gluonic field configurations with this property are the center vortices.  

   The center group is also singled out by the deconfinement phase
transition, which involves the breaking of a global $Z_N$ symmetry,
and certain features found in the deconfined phase are elegantly
explained in terms of the vortex picture.~\cite{T} Note that only
global symmetries can actually break spontaneously.  In the absence of
gauge fixing, the VEV of a Higgs field in any gauge theory, dual or
otherwise, is zero in any phase.  This is due to the Elitzur theorem,
and also to the status of local gauge symmetry as a genuine redundancy
in field variables.

   Let us consider the non-confinement of abelian electric charge in
more detail (cf.\ Ambj{\o}rn et al.~\cite{j3} for an extended
discussion).  Fixing to maximal abelian gauge, we write the link
variables in the usual form $U_\m = W_\m A_\m$ where $A_\m$ is the
abelian (diagonal) link variable.  Imagine integrating out the $W$ and
ghost fields, to obtain an effective abelian action
\beq
\exp\Bigl[S_{eff}[A]\Bigr] = \int DW_\m D(\mbox{ghosts}) ~ e^{S+S_{gf}}
\eeq
The reduction to $U(1)^{N-1}$ degrees of freedom is not particularly
significant in itself; this procedure could be
carried out for any subgroup of SU(N), including the $Z_N$ center.
Moreover, such effective actions are likely to be extremely
complicated and non-local.  The reduction only becomes physically 
interesting if $S_{eff}[A]$ takes on some particularly simple form at
large scales, e.g.\ if $S_{eff}[A]$ describes a dual
superconductor and/or monopole Coulomb gas of some kind.  There exist,
in fact, some very concrete proposals along these lines.~\cite{Suzuki}  
One feature which is always found in such proposals is that \emph{all} 
multiples of abelian electric charge are confined by the dual Meissner 
effect.  If that is really so, then an immediate consequence is that 
abelian Polyakov lines corresponding to any integer 
multiple of the electric charge ought to vanish in the confined phase.  
In SU(2) lattice gauge theory, this means that
\beq
      P_q = \langle \mbox{Tr}[(A_0A_0...A_0)^q] \rangle_{S_{eff}}
                = 0 ~~~\mbox{all~} q
\eeq
\ni In addition, if $P_{Mq}$ denotes the ``monopole dominance''
approximation~\cite{Suz1} to $P_q$, and if $S_{eff}[A]$ is well described by a 
monopole Coulomb gas, we would also expect 
\beq
      P_{Mq} \approx P_q    
\eeq
The fact is, however, that while 
both of the above relations hold at $q=1$ ($Z_2$ charged),~\cite{Suz1}
neither relation holds at $q=2$ ($Z_2$ neutral).~\cite{j3} 
The relevant $q=2$ Polyakov lines in the confined phase (for $T=4$
lattice spacings in the time direction), are shown in fig.\ \ref{P2T4} below. 
From the data showing $P_2 \ne 0$, we conclude that
there is no confinement for $q=2$.
From $|P_2| \gg |P_{M2}|$, it seems there is no monopole dominance either.
And from the fact that $P_2<0$ we find that positivity is broken as
well.  Actually, positivity can be restored by
going to spacelike maximal abelian gauge, but then $90^\circ$ rotation
symmetry is lost, and $q=2$ string-breaking still occurs.~\cite{j3}
\begin{figure}[h!]
\centerline{\psfig{figure=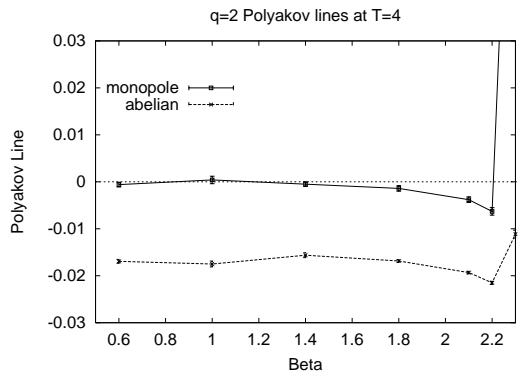,height=2.0in}}
\caption{Double-charge abelian Polyakov lines}
\label{P2T4}
\end{figure}

  Non-confinement of $q=2$ charge, and the breakdown of monopole
dominance, pose severe difficulties for 
dual-superconductor/dual abelian Higgs/ monopole Coulomb gas models
based on the abelian projection.  All of these pictures predict
confinement of all $q$; therefore none of them is a good description
of $S_{eff}[A]$.

  A similar objection can be raised to the monopole confinement picture
in the D=3 Georgi-Glashow model.  Although the monopole Coulomb gas picture,
developed by Polyakov, is certainly valid for some intermediate range of 
distances, this picture must break down asymptotically.  The reason is that in
a monopole Coulomb gas we find string tensions $\s_q \propto q$
between objects with $q$ units of U(1) charge.~\cite{j2}  But the 
Georgi-Glashow model has W-bosons with $q=2$.  Taking charge screening by 
these fields into account, we must get eventually
\beq
        \s_q = \left\{ \begin{array}{cl}
                  \s_1 & \mbox{odd} ~~ q \cr
                    0  & \mbox{even} ~ q \end{array} \right.
\eeq
which contradicts the Coulomb gas picture. There is a moral here: In a 
confining theory, \emph{massive charged fields are relevant to far-infrared 
vacuum structure,} and cannot be ignored.  

   These comments apply also to the Seiberg-Witten model.  The low-energy 
effective action, derived in this model, again explicitly
neglects the massive W-particles, and therefore misses the screening
effects due to those particles.  For this reason, the Seiberg-Witten 
effective action (which assumes locality)  is not really the same thing 
as a Wilsonian effective action obtained from integrating out massive 
charged fields, and does not adequately describe physics beyond the
double-charge screening scale.~\cite{j3}

   In pure SU(2) gauge theory in a physical abelian gauge,
non-confinement of $q=$ even charge can likewise be deduced from the
inevitable electric charge screening by off-diagonal gluons (the
W-fields of eq.\ (1)).  The effect is no mystery, but the consequences
are important.  While non-confinement ($\s_q=0$) for $q=$ even must be
a property of the true effective action $S_{eff}[A]$ for the abelian
field, a very different $q$-dependence ($\s_q \sim q$) is
found in dual superconductor and monopole gas models.  This indicates
that the latter are either incorrect or, at best, incomplete in some
way.  In contrast, the correct $q$-dependence of the abelian string
tension is quite natural in the framework of the vortex theory
(cf. Ambj{\o}rn et al.~\cite{j3}).


\section{Center Vortices at Strong Couplings}

  In strong-coupling lattice gauge theory in $D>2$ dimensions, we have both 
confinement for N-ality $\ne 0$ charges,
and color screening for N-ality $=0$ charges.
These facts suggest the existence of a vortex mechanism.
On the other hand, there is a bit of folklore about strong coupling,
namely, that confinement in $D>2$ dimensions is just
due to plaquette disorder, as in $D=2$ dimensions.  If so, vortices
(and any other topological objects), have nothing to do with confinement at
strong coupling.  This folklore, however, is misleading.

   Consider SU(2) lattice gauge theory at strong-coupling, and denote 
by $U(C)$ the product of link variables around loop $C$.  Let the minimal
area of a planar loop be decomposed into a set of smaller areas, bounded
by loops $\{C_i\}$.  We ask: Do the $\{U(C_i)\}$ fluctuate (nearly) 
independently, for large areas and small $\beta$?  The test is
whether
\beq
      <\prod_i F[U(C_i)]> \stackrel{?}{=}  \prod_i <F[U(C_i)]> 
\label{test}
\eeq
for any class function
\beq
        F[g] = \sum_{j\ne 0} f_j \chi_j[g]
\eeq
In fact, in $D=2$ dimensions, it is easy to show that this equality
is satisfied exactly.  However, for dimensions $D>2$, evaluating the
left- and right-hand sides of \rf{test} we find for the exponential
falloff on each side~\cite{j2}
\beq
  e^{-4\s P(C)} \prod_i {1\over 3} f_1 \gg \prod_i f_1 e^{-4\s P(C_i)}
\eeq
where the inequality holds for perimeters $P(C)\ll \sum_i P(C_i)$.
The conclusion is that the holonomies $U(C_i)$ do \emph{not} fluctuate 
independently, even at strong-coupling, for $D>2$. Where, then, does the 
area-law falloff come from?

   The question is resolved by extracting a center element from the 
holonomies
\beq
        z[U(C)] = \mbox{signTr}[U(C)] \in Z_2
\eeq
and asking if the center elements fluctuate independently; i.e
\beq
      <\prod_i z[U(C_i)]> \stackrel{?}{=}  \prod_i <z[U(C_i)]> 
\eeq
In fact, it is easy to show that they do:
\beq
  e^{-\s A(C)} \prod_i {3\over 4\pi} = \prod_i {3\over 4\pi} e^{-\s A(C_i)}
\eeq

   Thus, confining disorder is center disorder, at least
at strong couplings.~\cite{j2}  Confining configurations must disorder 
the center elements $z$, but not the coset elements, of SU(2) holonomies
$U(C_i)$.  Again, the only configurations known to have this 
property are center vortices.

   If center vortices are, in fact, the confining configurations of
strong-coupling lattice gauge theory, then it would be interesting if
this fact could be demonstrated analytically.  A reasonable conjecture
is that if the Wilsonian effective action could be computed at a scale
beyond the vortex thickness (4-5 lattice spacings at strong couplings),
then at this scale ``thin'' center vortices will be stable saddlepoints
of the action. 

  Suppose we define an effective long-range action $ S_{eff}$ by, e.g.
\beq
  \exp\Bigl[S_{eff}[V]\Bigr] = \int DU \prod_{l'} 
\d\Bigl[V^\dg_{l'}(UU..U)_{l'}-I\Bigr] e^{S_W[U]}
\eeq
where the V-lattice spacing is $L$ U-lattice spacings.  In D=2 dimensions
\bea
 \lefteqn{\exp\Bigl[S_{eff}[V]\Bigr]}
\non \\
       & =&  {\cal N}
\exp\left[ \sum_{P'}\log\left( 1 + \sum_{j=\oh,1,{3\over 2}} (2j+1)
    \left({I_{2j+1}(\b)\over I_1(\b)}\right)^{L^2} \chi_j[V(P')] \right) 
        \right]
\eea
But this must be wrong for $D>2$ dimensions, because it
leads to a perimeter-law falloff 
\beq
     \langle \chi_1[V(C)] \rangle \sim \exp[-\m \P(C)]
\eeq
with an $L-$dependent ``gluelump'' mass
\beq
      \m = 4L \log\left( {\b\over 4} \right) ~~~~~~~~ \mbox{(wrong)}
\eeq
The correct coefficient is
\beq
      \m = 4 \log\left( {\b\over 4} \right) 
\eeq   

   In fact, $S_{eff}$ in $D>2$ dimensions is non-local.  It contains
loops in all $j=$ integer representations, with perimeter-law
weightings, derived from diagrams in which plaquettes on the U-lattice
form a ``tube'' around a contour $C$ on the V-lattice.  These diagrams
lead to non-local contributions to $S_{eff}$ such as
\beq
  S_{eff}[V] \supset 
\left( {\b\over 4}\right)^{4(\P(C)-4)} \chi_1[V(C)]
\eeq
\begin{figure}[h!]
\centerline{\psfig{figure=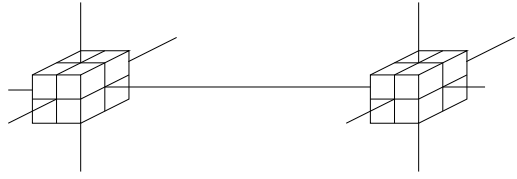,height=0.9in}}
\caption{Unintegrated links on the ``coarse'' V-lattice.}
\label{fig1}
\end{figure}

   We would like to derive a local effective action which would produce
at least the \emph{leading} contribution
to any Wilson loop on the V-lattice. To achieve this, we 
integrate over all links on the U-lattice
except on 2-cubes surrounding V-lattice sites, as shown in 
fig.\ \ref{fig1}.\footnote{We work in D=3 dimensions.  The extension
to D=4 should be straightforward.}
This defines an effective action $\tS_L$
\bea
      Z  & =& \int DV \int \prod_{l\in 2-cubes} d\tU_l 
\non \\
   & &  \left\{ \int \prod_{l''\not\in 2-cubes} dU_{l''} 
 \prod_{l'} \d\Bigl[V^\dg_{l'}(UU..U)_{l'}-I\Bigr] e^{S_W[U]} \right\}
\non \\
        &=& \int DV \int \prod_{l\in 2-cubes} d\tU_l ~ 
                      \exp\Bigl[\tS_L[V,\tU]\Bigr]
\non
\eea
Introduce group-valued plaquette variables in the 2-cubes $h,g$, where
$h$ variables run around plaquettes on the surface of the 2-cube, and
$g$ variables run around plaquettes in the interior.  Both sets of contours
begin and end at the center of the 2-cube. After a number of 
manipulations, which include changing variables from $\tU$ to $h,g$
and integrating over the $g$ variables, we obtain
\bea
   Z &\approx&  \int DV Dh \prod_{2-cubes~K}~ 
\left\{ 1 + 2\left({\b\over 4}\right)^3 \sum_{c\in K} 
\chi_\oh[(hhh)_c] \right.
\non \\
      & &\left.  + 2\left({\b\over 4}\right)^4 
                 \sum_{\stackrel{adjacent}{c_1c_2\in K}} 
                  \chi_\oh[(hhh)_{c_1}(hhh)_{c_2}] + ... \right\}
\non \\
  &\times& \exp\left[ {\b\over 2} \sum \mbox{Tr}[h] 
    + 2\left(\b\over 4\right)^{4(L-2)}\sum_{l'}
  f_{l'}^{ijkl} \mbox{Tr}[h^\dg_{ij}V_{l'}
            h^\dg_{kl}V^\dg_{l'}] \right.
\non \\
  & & \left.  + 2\left({\b\over 4}\right)^{L^2} \sum_{P'}
              \mbox{Tr}[V V V^\dg  V^\dg ] \right]
\eea
where $f_{l'}^{ijkl}=1$ if two plaquettes on neighboring 2-cubes can
be joined by a cylinder of plaquettes on the U-lattice adjacent to 
V-link $l'$; $f_{l'}^{ijkl}=0$
otherwise.  This resembles an adjoint-Higgs theory, with an SU(2)
gauge field $V_\m$ coupled to 24 ``matter'' fields $h$ in the adjoint
representation.  Note that for large $L$, the ``Higgs'' potential term
is much larger than the ``kinetic'' (V-link) and pure-gauge
(V-plaquette) terms, so the $h$-fields fluctuate almost independent of
$V_\m$.  We then do a unitary gauge-fix of the $h$-fields (which
leaves a remnant $Z_2$ symmetry), and integrate out the remaining $h$
d.o.f.\ to obtain
\bea
      S_{eff}[V] &\approx& 
S_{link}[V,\langle h \rangle_h] + S_{plaq}[V]
\non \\
        & =&  2\left({\b\over 4}\right)^{4(L-2)}
           \sum_{l'} f_{l'}^{ijkl}\mbox{Tr} \Bigl[
      \langle h^\dg_{ij} \rangle_h V_{l'} 
      \langle h^\dg_{kl} \rangle_h V^\dg_{l'} \Bigr]
\non \\
       & &  + 2\left({\b\over 4}\right)^{L^2}
             \sum_{P'} \mbox{Tr}[VVV^\dg V^\dg]
\eea

  Now look for saddlepoints.  We find that $ S_{link}$ is maximized
at
\bea
         V_\m(\vec{n}) & =&  Z_\m(\vec{n}) 
                       \times g(\vec{n}) g^\dg(\vec{n}+\hat{\m})
\non \\
   Z_\m    & =&  \pm 1
\non
\eea
where $g(\vec{n}) g^\dg(\vec{n}+\m)$ is fixed by the particular 
unitary gauge
choice, while $ S_{plaq}$ is maximized if $ZZZZ=+1$.  This is the
unitary gauge ground state.
Create a thin center vortex on this state by a discontinuous gauge
transformation, e.g.\

\bea
      Z_y(\vec{n}) & =&  \left\{ \begin{array}{rl}
                    -1 & n_1 \ge 2, ~ n_2=1 \cr 
                    +1 & \mbox{otherwise} \end{array} \right.
\non \\
     Z_x(\vec{n}) & =&  Z_z(\vec{n}) = 1 
\non
\eea
This configuration is stationary:  $ S_{link}[V]$ is
still a maximum, and $S_{plaq}$ is extremal (max or min) on all plaquettes.
Stability depends on the eigenvalues of
\beq
     {\d^2 S_{eff} \over \d V_\m(n_1) \d V_\n(n_2)} =
        {\d^2 S_{link} \over \d V_\m(n_1) \d V_\n(n_2)}
      + {\d^2 S_{plaq}  \over \d V_\m(n_1) \d V_\n(n_2)} 
\eeq
and we find
\bea
      {\d^2 S_{link}  \over  \d V_\m(n_1) \d V_\n(n_2)}
         &\sim& \left({\b\over 4}\right)^{4(L-2)+12}
\non \\
      {\d^2 S_{plaq}  \over  \d V_\m(n_1) \d V_\n(n_2)}
         &\sim&  \left({\b\over 4}\right)^{L^2}
\eea

  The crucial observation is that for $\b/4 \ll 1$ and
\beq
      4(L-2) + 12 < L^2  ~~~\Longrightarrow~~~ L \ge 5
\eeq
the contribution of $ \d^2 S_{plaq}/\d V \d V$ to the stability 
matrix (and therefore to the eigenvalues of the stability matrix) 
is negligible compared to
$ \d^2 S_{link}/\d V \d V$, which has only stable modes.  This implies:
\begin{enumerate}
\item \emph{Vortex Stability:} The thin vortex is a stable saddlepoint 
of the full effective action $S_{eff}$ at $L\ge 5$.
\item \emph{Vortex Thickness:}  A ``thin'' vortex on the V-lattice means 
thickness $< L$ on the U-lattice.  This means that stable 
center vortices are $\approx 4-5$ lattice spacings thick.  For the strong 
coupling Wilson action, this is the distance where the adjoint string breaks!
The correspondence between the adjoint string-breaking length, and
the thickness of center vortices, has been emphasized by our group 
in connection with Casimir scaling~\cite{Cas} (see also 
Cornwall~\cite{Corn}).  
\item \emph{Percolation:}  From $S_{eff}$, we see that center vortices 
in D=3 cost an action $8 (\b / 4)^{L^2}$/unit length,
while the entropy is O(1)/unit length.  Since entropy $\gg$ action, this 
implies that vortices percolate through the lattice, and confine N-ality 
$\ne 0$ charge.
\end{enumerate}

\section{P-Vortices, Gauge Copies, and Lattice Size}

   The original calculations of center-projected Creutz ratios 
$\chi_{cp}(I,I)$, in direct maximal center gauge, used 3 gauge copies 
for gauge-fixing each lattice (picking the best of the three).~\cite{Jan98}   
Very recently Bornyakov et al.~\cite{borny} 
have claimed that $\chi_{cp}(I,I)$ varies with the number of
gauge copies used, and disagrees, in the large copy number limit, with
the unprojected string tension by as much as 30\%.  

   In our opinion, the reported strong disagreement between projected
and unprojected string tensions is due to finite-size effects. Lattice
sizes used by Bornyakov et al.\ were $12^4$ at $\b=2.3,2.4$, and
$16^4$ at $\b=2.5$,~\cite{borny} while our published results were
obtained on $16^4$ lattices at $\b=2.3,2.4$ and $22^4$ lattices at
$\b=2.5$.~\cite{Jan98} Projected lattices are more sensitive to
finite-size effects than unprojected lattices, and this is probably
due to the fact that center projection has difficulty finding
vortices, when most of the lattice volume is taken up by the vortex
cores.~\cite{vf} There are now good estimates for vortex thickness,
coming from three sources: First, from the ratio of ``vortex-limited''
Wilson loops.~\cite{Jan98} Second, from the adjoint string-breaking
distance, measured by de Forcrand and Philipsen.~\cite{dFP} Third, the
vortex thickness is found in a very interesting calculation, reported
here by Terry Tomboulis, of the vortex free energy vs.\ lattice
size.~\cite{KT} All three estimates are in rough agreement, and give a
vortex thickness of a little over one fermi.  This means that center
vortices are $\approx 12-14$ lattice spacings thick at $\b=2.5$; a
$16^4$ lattice may just be too small on this scale.

\begin{figure}[h!]
\centerline{\psfig{figure=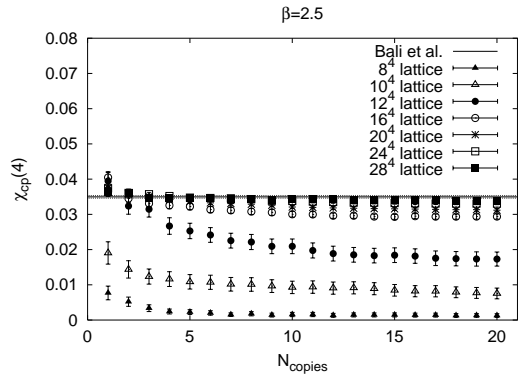,height=2.0in}}
\caption{Finite size and gauge copy dependence of the projected Creutz
ratio $\chi_{cp}(4,4)$ at $\b=2.5$. The solid line is the full SU(2) string
tension.}
\label{aborny}
\end{figure}

   We have therefore repeated the calculation of Bornyakov et al.\ on
a variety of lattice sizes.  Some typical results at $\b=2.5$ are shown
in fig.\ \ref{aborny}, where the finite size dependence is clearly seen.
When the lattice volume is large enough, increasing the number of copies
does not seem to make any substantial difference to our previously 
reported results for projected Creutz ratios and vortex densities.
The details will be presented in a separate publication.~\cite{coming}

   Finally, some questions about propagating ghosts in center gauges
were raised at this meeting in the summary talk by
Schierholz.~\cite{GS} Center gauges, like Landau gauge, are not
ghost-free, and this would be a real problem if the aim of center
gauge-fixing were to eliminate all unphysical modes in the
Lagrangian.~\cite{tH} But the issue has little relevance, in our
opinion, to the actual purpose of center gauge fixing, which is used
in conjunction with center projection as a vortex finder.  The
rationale underlying this procedure,~\cite{vf} and its empirical
success in finding vortices on thermalized
lattices,~\cite{Jan98,dFE,mog} have been discussed at length
elsewhere. The same speaker questions whether vortex physics will be
found to be consistent with instanton physics.  We see no evidence of
a problem in this area; in fact there are some suggestive findings to
the effect that removing vortices from a lattice configuration also
removes the topological charge.~\cite{dFE}  In any case, the study of
instanton physics in relation to center vortices has only just
begun,~\cite{Negele,Engel} and there is no reasonable basis, at this
stage, for strong conclusions.

    In this talk I have indicated how the center vortex picture of
confinement can be derived at strong lattice couplings, and why this
picture is attractive at any gauge coupling. The strong-coupling analysis
shows that center vortices are stabilized by color-screening terms in
the long-range effective action, and that screening terms 
dominate the action at the 
adjoint string-breaking scale.  We expect that these qualitative features 
of the strongly coupled gauge theory are also found in the continuum 
$\b\ra \infty$ limit.

\section*{Acknowledgments}

   I would like to thank Hideo Suganuma, Hiroshi Toki, and the other 
organizers of Confinement 2000 for inviting me to participate in this 
stimulating meeting. This work was supported by the US Dept.\
of Energy under Grant No.\ DE-FG03-92ER40711.

\end{document}